\documentclass[twocolumn,superscriptaddress,amsfont,amssymb,amsmath, showpacs,balancelastpage, nofootinbib]{revtex4-1}
\usepackage{graphicx,longtable,mathrsfs,color,array}
\usepackage[hidelinks]{hyperref}
\usepackage[usenames,dvipsnames]{xcolor}
\usepackage{amssymb,amsmath,mathtools,mathrsfs}
\usepackage{epsfig,subfigure,placeins,float}
\usepackage{booktabs,longtable,ctable,multirow}
\usepackage{exscale,relsize}
\usepackage[normalem]{ulem}
\usepackage{enumerate}
\usepackage{times, mathptmx}
\usepackage{colortbl} 
\usepackage[dvipsnames]{xcolor} 
\usepackage{pifont} 
\usepackage{amsmath} 
\usepackage{adjustbox}

\newcommand{\be}{\begin{equation}}
\newcommand{\ee}{\end{equation}}

\newcommand{\gammav}{v}
\newcommand{\Upsour}{\gamma}

\newcommand\alphaB{\alpha_{\text{B}}}

\newcommand\alphaK{\alpha_{\text{K}}}

\newcommand\alphaH{\alpha_{\text{H}}}
\newcommand\alphaV{\alpha_{\text{V}}}
\def\d{\delta}

\allowdisplaybreaks

\begin{document}

\title{{Vainshtein Screening in Scalar-Tensor Theories before and after GW170817:\\
Constraints on Theories beyond Horndeski}}
\author{Alexandru Dima}
\affiliation{SISSA -- International School for Advanced Studies, Via Bonomea 265, 34136 Trieste, Italy}
\author{Filippo Vernizzi}
\affiliation{Institut de physique th\' eorique, Universit\'e  Paris Saclay 
CEA, CNRS, 91191 Gif-sur-Yvette, France}
        \date{\today}

\begin{abstract}

Screening mechanisms are essential features of dark energy models mediating a fifth force on large scales. 
We study the regime of strong scalar field nonlinearities, known as Vainshtein screening, in the most general scalar-tensor theories propagating a single scalar degree of freedom. 
We first develop an effective approach to parameterize cosmological perturbations beyond linear order for these theories. In  the quasi-static  limit,  the fully nonlinear effective Lagrangian contains six independent  terms, one of which starts at cubic order in perturbations. 
We compute the  two gravitational potentials around a spherical body. 
Outside and near the body,  screening  reproduces standard gravity, with a modified gravitational coupling. Inside the body, the two potentials are different and depend on the density  profile, signalling the breaking of the Vainshtein screening.  
We provide the most general expressions for these modifications, revising and extending  previous results.
We apply our findings to show that the combination of the GW170817  event, the Hulse-Taylor pulsar and stellar structure physics,  constrain the parameters of these general theories at the level of $  10^{-1}$, and of GLPV theories at the level of $10^{-2}$.

\end{abstract}

\maketitle

\section{Introduction} 

The recent simultaneous observation of gravitational waves and gamma ray bursts  from GW170817 \cite{TheLIGOScientific:2017qsa} and GRB 170817A  \cite{Goldstein:2017mmi} has allowed to constrain very precisely the relative speed between gravitons and photons. This measurement has had dramatic impact on the parameter space of modified gravity theories characterized by a single scalar degree of freedom \cite{Creminelli:2017sry,Sakstein:2017xjx,Ezquiaga:2017ekz,Baker:2017hug} (see \cite{Lombriser:2015sxa,Bettoni:2016mij} for earlier work). In particular,  the so-called Horndeski theories \cite{Horndeski:1974wa,Deffayet:2011gz}, a class of well-studied scalar-tensor theories that are often used as benchmarks to parameterize modifications of gravity, have been drastically simplified. Their higher-order Lagrangian terms, quadratic and cubic in  second derivatives of the field, predict a speed of gravitational waves that differ from that of light and are thus ruled out.
This fact has triggered renewed interest for the surviving theories, i.e.~those extending the Horndeski class that are  compatible with the GW170817 observation, such as certain subclasses of  Gleyzes-Langlois-Piazza-Vernizzi (GLPV) theories \cite{Gleyzes:2014dya,Gleyzes:2014qga}. 

Lagrangian terms with higher-derivatives are crucial  
to suppress, via the so-called Vainshtein  mechanism \cite{Vainshtein:1972sx,Babichev:2013usa}, the fifth force exchanged by the scalar and responsible for the modifications of gravity on large scale. On the other hand, theories  extending the Horndeski class are known to display a breaking of the Vainshtein screening inside matter \cite{Kobayashi:2014ida}, a phenomenon that  has allowed to constrain the parameter space of these theories with astrophysical observations \cite{Koyama:2015oma,Saito:2015fza,Sakstein:2016ggl,Babichev:2016jom,Sakstein:2016oel}.

The purpose of this paper is to study the Vainshtein mechanism in the general framework of the degenerate theories introduced in \cite{Crisostomi:2016czh,Langlois:2015cwa}, which includes the Horndeski class and theories beyond Horndeski  \cite{Zumalacarregui:2013pma,Gleyzes:2014dya,Gleyzes:2014qga}.
We will consider only theories that can be related to the Horndeski class by an invertible metric redefinition \cite{Crisostomi:2016czh,Achour:2016rkg}. In the classification of Ref.~\cite{Langlois:2015cwa} (or \cite{Crisostomi:2016czh}), they are called degenerate higher-order scalar-tensor theories of  class Ia (or extended scalar-tensor theories of class N-I). 
Moreover, we will focus on theories up to {\em quadratic} in the second derivative of the scalar field. In particular, we do not consider theories {\em cubic} in the second derivative of the scalar \cite{BenAchour:2016fzp}, whose Vainshtein mechanism has been poorly studied due to its complexity, because they are anyway ruled out by the observation of GW170817 \cite{Creminelli:2017sry,Ezquiaga:2017ekz,Baker:2017hug}.

We do so by reducing these theories to their essential elements with the use of the Effective Field Theory (EFT) of dark energy description developed in \cite{Creminelli:2008wc,Gubitosi:2012hu,Gleyzes:2013ooa,Gleyzes:2014rba,Langlois:2017mxy,Cusin:2017mzw}.
Moreover, we focus on scales much smaller than the Hubble radius and we restrict to non-relativistic sources, in which case scalar fluctuations satisfy the quasi-static approximation. 
We will first derive very general expressions for the two potentials in the Vainshtein regime. This will be important to extend and clarify previously obtained results. Then, restricting to theories propagating gravitons at the speed of light, we will use our expressions to show that a combination of constraints from stellar structure \cite{Sakstein:2015zoa,Saito:2015fza} and from precise measurements of the decrease of the orbital period in the Hulse-Taylor binary pulsar  severely constrain these scenarios.

During the preparation of this work, Refs.~\cite{Crisostomi:2017lbg,Langlois:2017dyl} have appeared, where some of the results derived in this article are independently obtained   using different approaches.

\section{Degenerate Higher-Order Scalar-Tensor theories}
Let us consider a scalar-tensor field theory described by an action including all possible quadratic combinations up to second derivatives of the field $\phi$  \cite{Langlois:2015cwa},
\be
\begin{split}
\label{TypeIa}
S = \int & d^4 x {\cal L} = \int d^4 x \sqrt{-g} \Big[ P(\phi,X)+ Q(\phi,X) \Box \phi \\ 
&+ f(\phi,X) {}^{(4)}\!R + \sum_{I=1}^5 a_I (\phi,  X ) L_I (\phi, \phi_{;\nu},  \phi_{;\rho \sigma})  \Big] \;,
\end{split}
\ee
where ${}^{(4)}\!R$ is the 4D Ricci scalar.
A semicolon denotes the covariant derivation, $X \equiv - \phi_{;\mu} \phi^{;\mu}/2 $, and the $L_I$ are defined by
\be
\begin{split}
L_1&=\phi_{;\mu\nu}\phi^{;\mu\nu} \;, \quad  L_2=(\phi^{;\mu}_{;\mu})^2 \;, \quad  L_3=(\phi^{;\mu}_{;\mu})(\phi^{;\rho} \phi_{;\rho\sigma}\phi^{;\sigma})\;, \\
  L_4&=\phi^{;\mu}\phi_{;\mu\nu}\phi^{;\nu\rho}\phi_{;\rho} \;, \quad  L_5=(\phi^{;\rho} \phi_{;\rho\sigma}\phi^{;\sigma})^2 \;.
\end{split}
\ee
In the following, we are going to focus on Type Ia theories, which  satisfy $a_1+a_2 =0$ and two other degeneracy conditions \cite{Langlois:2015cwa} that fix two  functions, for instance $a_4$ and $a_5$. Degenerate theories that are not in this class have been shown to propagate scalar fluctuations with  sound speed squared with opposite sign to the sound speed squared of tensor fluctuations \cite{Langlois:2017mxy} and we will not consider them here.
The theory with $a_3+a_4=a_5=0$ is degenerate also in the absence of gravity. In particular, 
in the notation of \cite{Gleyzes:2014dya} 
this includes the case $f=G_4$, $a_1 = -a_2 = -G_{4,X}$ (a comma denotes the derivative with respect to the argument) and $a_3=a_4=a_5=0$, corresponding to quartic Horndeski theories, and the case $f=G_4$, $a_1 = -a_2 = -  G_{4,X} +2 XF_4$, $a_3=-a_4= - 2 F_4$ and $a_5=0$, corresponding to quartic GLPV theories. The functions $P$ and $Q$ do not affect the degeneracy character of the theory.

\section{Effective Theory of Dark Energy}
To describe cosmological perturbations around a FRW solution in theories with a preferred slicing induced by a time-dependent  scalar field, it is convenient to use the EFT of dark energy.
To formulate the action \eqref{TypeIa} with the conditions $a_1+a_2=0$, we use the ADM metric decomposition, where the line element reads $ds^2 = - N^2 dt^2 + h_{ij} (dx^i + N^i dt) (dx^j + N^j dt)$, and we choose the time as to coincide with the uniform field hypersurfaces. Moreover, we are going to focus only on the operators that contribute in the {\em quasi-static} limit.

In this gauge, expanded around a flat FRW background $ds^2=-dt^2 + a^2(t) d\vec x^2$, the full nonlinear action reads
\be
\begin{split}
\label{EFTaction}
&S_{\rm QS} = \int  d^4 x \sqrt{h}  \frac{M^2}{2} \big( -(1+\delta N)\delta {\cal K}_2 +c_T^2 {}^{(3)}\!R +4 H \alphaB \delta K \delta N \\
 &  +(1+\alphaH) {}^{(3)}\!R \delta N 
+4 \beta_1 \delta K V + \beta_2 V^2 + {\beta_3} a_i a^i
+ \alphaV \delta N \delta {\cal K}_2 \big)\;.
\end{split}
\ee
Here $H\equiv \dot a/a$ (a dot denotes the time derivative), $\delta N \equiv N-1$, $\d K_i^j \equiv K_i^j- H \delta_i^j$  is the perturbation of the extrinsic curvature of the time hypersurfaces, $\delta K$ its trace and ${}^{(3)}\!R$ is the 3D Ricci scalar of these hypersurfaces. Moreover, $\delta {\cal K}_2 \equiv \delta K^2 - \delta K_{i}^j \delta K^{i}_{j}$, $V \equiv (\dot N - N^i \partial_i N)/N$ and $a_i \equiv \partial_i N/N$.

We have also defined the effective Planck mass, which normalizes the graviton kinetic energy,  by $M^2 \equiv 2( f - 2 a_2 X)$ and a few independent parameters, related to the functions in \eqref{TypeIa} by
\newcommand{\bun}{\beta_1}
\newcommand{\bdeux}{\beta_2}
\newcommand{\btrois}{\beta_3}
\be
\label{alphabeta}
\begin{split}
\alphaB & = \alphaV -3 \beta_1 + \dot \phi ( f_\phi + 2 X f_{,\phi X} + X Q_{,X}) /(M^2H) \;, \\
c_T^2 &= {2 f}/{M^2}\,, \qquad   \alphaH= 4 X (a_2 - f_{,X})/M^2  \,,
\\
\bun& =  2 {X} (f_{,X}- a_2+ a_3 X)/{M^2} \,, \\
\alphaV & = {4 X ( f_{,X} - 2a_2 -2X a_{2,X} )}/{M^2} \;.
\end{split}
\ee
The function $c_T^2$ is the fractional difference between the speed of gravitons and photons. Sometimes called braiding \cite{Bellini:2014fua}, the function $\alphaB$  measures the kinetic mixing between metric and scalar fluctuations \cite{Creminelli:2008wc}. The function $\alphaH$ measures the kinetic mixing between matter and the scalar in GLPV theories \cite{Gleyzes:2014dya,Gleyzes:2014qga,DAmico:2016ntq} and vanishes for Horndeski theories. The functions $\beta_1$, $\beta_2$ and $\beta_3$ parameterize the presence of higher-order operators.
In the EFT of dark energy formulation, the  degeneracy conditions that ensure that the action \eqref{EFTaction} describes the propagation of a single scalar degree of freedom are  \cite{Langlois:2017mxy}
\be
\label{degeneracy}
\bdeux=-6\bun^2\,,\qquad   \btrois=-2\bun\left[2(1+\alphaH)+\bun c_T^2 \right] \;,
\ee
so that we do not need the explicit expression for $\beta_2$ and $\beta_3$ in terms of the functions defining \eqref{TypeIa}.
We will impose these conditions later.
Finally, the operator proportional to $\alphaV$ is the only one that starts cubic in the perturbations.  
In the nonlinear EFT action, it was introduced (as $-\alpha_{\rm V1}$) in \cite{Cusin:2017mzw} to describe  nonlinear dark energy perturbations. 
Notice that the action \eqref{EFTaction} does not include the kineticity \cite{Bellini:2014fua} Lagrangian term $\alphaK \delta N^2$, because it can be neglected in the quasi-static limit \cite{Cusin:2017mzw}. 
The total number of independent parameters, and thus of Lagrangian operators, is thus   six.

\renewcommand{\arraystretch}{1.2}
\begin{table}
\begin{center}
\begin{adjustbox}{max width=\textwidth}
  \begin{tabular}{ | l | |c | c | c | c |  c |  c | }
    \hline   & $ M^2 $   & $ \alphaB $   & $ c_T^2-1 $   &  $\alphaH$ &  $\beta_1$ & $\alphaV$   \\  
    [0.0cm]  \hline\hline $P(\phi,X) $  & $0$ & $0$ & $0$ & $0$ & $0 $ & $0$  \\ 
        [0.0cm] \hline $Q(\phi,X) \Box \phi$  & $0$ & $\checkmark$ & $0$ & $0$ & $0 $ & $0$  \\ 
        [0.0cm] \hline Quartic Horndeski  & $\checkmark$ & $\checkmark$ & $\checkmark$ & $0$ & $0 $ & $\checkmark$  \\ 
    [0.0cm] \hline Quartic  GLPV      & $\checkmark$ & $\checkmark$ & $\checkmark$ &  $\checkmark$ & $0$ & $\checkmark$  \\ 
    [0.0cm] \hline  Quadratic DHOST   & $\checkmark$ & $\checkmark$  & $\checkmark$ & $\checkmark$ & $\checkmark$  & $\checkmark$ \\    
        [0.0cm] \hline   After GW170817  & free & free  & 0 & free & free  & $-\alphaH$ \\    
        [0.0cm]  \hline
  \end{tabular}
\end{adjustbox}
\end{center}
\caption{Lagrangian operators of the EFT of dark energy  allowed in various theories and the  consequences of  the equality between speed of gravity and light on these theories. }
  \label{table}
\end{table}
We summarize the relation between the EFT operators and the corresponding covariant Lagrangians in Table~\ref{table}, where we also state in which way  these operators are affected by the equality between the speed of gravity and light, see \cite{Creminelli:2017sry} and discussion below.

\section{Action in Newtonian gauge}  

We now expand the Lagrangian \eqref{TypeIa} around an FRW background. We consider only scalar fluctuations  in the Newtonian gauge, where $\delta N= \Phi$, $h_{ij} = a(t)^2 (1-2 \Psi) \delta_{ij}$ and $N^i=0$. Without loss of generality, we take $\phi = t  + \pi (t, \vec x)$. 

In the quasi-static regime, time derivatives are of  order  Hubble  and the Lagrangian  is dominated by terms with $2 (n-1)$ spatial derivatives for $n$ fields. Considering only these terms, one obtains 
\newcommand{\AAone}{c_1 }
\newcommand{\AAtwo}{c_2 }
\newcommand{\AAthree}{c_3 }
\newcommand{\AAfour}{c_4 }
\newcommand{\AAfive}{c_5 }
\newcommand{\AAsix}{c_7 }
\newcommand{\AAseven}{c_6 }
\newcommand{\AAeight}{c_8 }
\newcommand{\AAnine}{c_9 }
\newcommand{\BBone}{b_1 }
\newcommand{\BBtwo}{b_2 }
\newcommand{\BBthree}{b_3 }
\newcommand{\BBfour}{b_4 }
\newcommand{\BBfive}{b_5 }
\newcommand{\BBsix}{b_6 }
\newcommand{\CCone}{d_1 }
\newcommand{\CCtwo}{d_2 }
\be
\label{Lagrangian}
\begin{split}
&S_{\rm QS} = \int  d^4 x    \frac{M^2a}2  \Big[     \big( \AAone \Phi  + \AAtwo  \Psi  + \AAthree \pi \big)\nabla^2 \pi  + \AAfour \Psi \nabla^2 \Phi 
\\
&+ \AAfive \Psi \nabla^2 \Psi  + \AAseven   \Phi \nabla^2 \Phi + \big( \AAsix \dot \Psi  + \AAeight  \dot \Phi + \AAnine  \ddot \pi \big) \nabla^2 \pi  
\\
& + \frac{\BBone}{a^2}     {\cal L}_3^{\rm Gal} +  \frac{1}{a^2} \big(  \BBtwo  \Phi  + \BBthree  \Psi \big) {\cal E}_3^{\rm Gal}  + \frac{1}{a^2} \big( \BBfour  \nabla_i \Psi + \BBfive   \nabla_i \Phi  \\
&+ \BBsix  \nabla_i \dot \pi  \big) \nabla_j \pi \Pi_{ij}     + \frac{1}{a^4} \big(   \CCone    {\cal L}_4^{\rm Gal} + \CCtwo   \nabla_i \pi \nabla_j \pi \Pi^2_{ij}  \big) \Big]\;.
\end{split}
\ee
Here, adopting the  notation of \cite{Kobayashi:2014ida}, 
$\Pi_{ij} \equiv \nabla_i \nabla_j \pi$, $\Pi^n_{ij} \equiv \nabla_i \nabla_{k_1} \pi \nabla_{k_1} \nabla_{k_2} \pi \ldots \nabla_{k_{n-1}} \nabla_j \pi $ and $ [\Pi^n] \equiv \delta^{ij} \Pi_{ij}^n$
 we have defined ${\cal L}_3^{\rm Gal}  \equiv  - \frac12 (\nabla \pi )^2 [ \Pi] $, ${\cal L}_4^{\rm Gal} \equiv  - \frac12  (\nabla \pi )^2 {\cal E}_3^{\rm Gal} $ and ${\cal E}_3^{\rm Gal}  \equiv  [\Pi]^2 - [\Pi^2]$. The coefficients $c_i$, $b_i$ and $d_i$ are time-dependent functions related to the functions $P$, $Q$, $f$, $a_I$  defining \eqref{TypeIa},  and their  derivatives, evaluated on the background solution.
 
 Equivalently, eq.~\eqref{Lagrangian} can  also be obtained  from the EFT action \eqref{EFTaction}, after introducing the scalar fluctuation $\pi$ by a time diffeomorphism $t \to t + \pi(t,\vec x)$ \cite{Cheung:2007st}. In this case, the coefficients $c_i$, $b_i$ and $d_i$ can be expressed in terms of the EFT parameters. The coefficients $c_{1}$, $c_2$, $c_3$ and $\BBone  $  are  functions of $M^2$, $\alphaB$, $c_T^2$, $\alpha_{\rm H}$, $\beta_1$, $\beta_3$ and $H$ (and their time derivatives) but we do not need their explicit expressions for the following discussion. The other coefficients are given by
\be
\label{Definitions}
\begin{split}
\AAfour & =  4  (1+ \alphaH)\;, \ \
\AAfive  =   -  2c_T^2  \;, \ \
\AAseven   =   -  \beta_3\;, \\
\AAsix & =   {4 \alphaH}   \;, \ \
\AAeight   =   - 2  {(2\beta_1 + \beta_3)}   \;, \ \
\AAnine   =     { 4\beta_1 + \beta_3}  \;, \\
\BBtwo  & =      \alphaV - \alphaH -  4 \beta_1  \;, \ \
\BBthree   = c_T^2-1  \;, \
\BBfour   =   - \AAsix \;, \\
\BBfive   & =  - \AAeight \;, \ \
\BBsix   =   - 2 \AAnine \;, \ \
\CCone    =  {- \BBthree-\BBtwo }  \;,  \ \
\CCtwo    =  \AAnine \;.
\end{split}
\ee
The relevant nonlinear couplings dominating in the Vainshtein regime will be the quartic ones in \eqref{Lagrangian}, i.e.~those proportional to $\CCone$ and $\CCtwo$.
Note  that they contain $c_T^2$, $\alphaH$, $\alphaV$, $\beta_1$ and $\beta_3$, but not $\alphaB$.  See more on this  below. 
For $ \beta_{1}=\beta_2=\beta_3= 0$, it is straightforward to verify that the above action agrees with those given in  \cite{Kimura:2011dc} for Horndeski and in  \cite{Kobayashi:2014ida} for GLPV theories.

To study the behaviour of $\Phi$, $\Psi$ and $\pi$ around dense matter sources,  we add to the action \eqref{Lagrangian} the coupling with non-relativistic matter with energy density $\rho_{\rm m} = \bar \rho_{\rm m}(t) + \delta \rho_{\rm m}(t,\vec x)$, i.e., 
\be
S_{\rm m} = - \int d^4 x a^3 \Phi \delta \rho_{\rm m}  \;.
\ee

\section{Vainshtein mechanism} 

To study the Vainshtein regime, we take matter to be described by some overdensity, spherically distributed around the origin. 
We define
\be
x \equiv \frac{1}{\Lambda^3} \frac{\pi'}{a^2 r} \;, \ \ y \equiv \frac{1}{\Lambda^3} \frac{\Phi'}{a^2 r} \;, \ \ z \equiv \frac{1}{\Lambda^3} \frac{\Psi'}{a^2 r} \;, \ \ {\cal A} \equiv \frac{1}{ 8 \pi M^2 \Lambda^3} \frac{m}{ r^3} \;,
\ee
where a prime denotes the derivative with respect to the radial distance $r$, $m (t,r)\equiv  4\pi \int_0^r \tilde r^2 \delta \rho_{\rm m} (t ,\tilde r) d \tilde r $ and $\Lambda$ is some mass scale of order $\Lambda \sim (M H^2)^{1/3}$.
Integrating   over space the equations obtained by varying the action \eqref{Lagrangian} respectively with respect to $\Phi$ and $\Psi$, 
and using Stokes theorem,  we obtain
\begin{align}
( \AAone -\dot \AAeight  - 3 H\AAeight )x+2\AAseven y+\AAfour z  -\AAeight  \dot{x} &  \nonumber
\\ + 2\Lambda ^3 x \left[ (2\BBtwo   -\BBfive  ) x- \BBfive  r x' \right]  & = 4  {\cal A} \;,  \label{VPhi}\\
(\AAtwo -\dot \AAsix - H \AAsix) x +\AAfour y+2\AAfive z -\AAsix \dot{x} & \nonumber
\\ + 2\Lambda^3x\left[ (2\BBthree  -\BBfour ) x- \BBfour  rx' \right] & =0  \label{VPsi} \;.
\end{align}
By applying the analogous procedure to the equation obtained by varying the action with respect to $\pi$, we get
\be
\label{xeq}
\begin{split}
& 2 \tilde \AAthree x+ \tilde \AAone y + \tilde \AAtwo z+ 2 \tilde \AAnine   \dot{x}   
+\AAeight  \dot{y} +\AAsix\dot{z}   + 2\AAnine  \ddot{x} \\ 
&+2 \Lambda^3 \big\{ 2 \tilde \BBone   x^2 +  (5H\BBsix +\dot \BBsix ) r x x'
+  \BBsix  ( 5 x \dot{x}+2  r x \dot{x}' + r\dot{x}x')  \\ 
&+x\big[ ( 4\BBtwo +3\BBfive  ) y+( 4\BBthree  +3\BBfour ) z  +\BBfive  ry' 
+\BBfour rz' \big]  \big\}  \\ 
&+ 8\Lambda ^6  \big\{  (\CCone    +3  \CCtwo  ) x^3 + \CCtwo   x \big[ r^2(x')^{2}+rx(6 x ' +rx'' ) \big] \big\}  =0 \;.
\end{split}
\ee
The coefficients with the tildes are related to those without the tildes and to their time derivatives, but we do not need their explicit expression for what follows.
Even though we have integrated  over space the equations obtained from the variation of the action \eqref{Lagrangian} and used $x$, $y$ and $z$, which have a derivative on the fields, these equations contain terms with two derivatives, indicating that they are higher than second order.

Equations \eqref{VPhi} and \eqref{VPsi}  are linear in $y$ and $z$, and can be solved for these two variables and their solutions can be replaced in eq.~\eqref{xeq}  to obtain an equation for $x$ only. Using the definitions \eqref{Definitions} for the time-dependent coefficients $c_i$, $b_i$ and $d_i$ and imposing the degeneracy conditions \eqref{degeneracy}, the space and time derivatives on $x$ cancel and one remains with
\be
\label{NLeq}
x^3 + \gammav_1 x^2 +  ( \gammav_2 + \gammav_3 {\cal A}+ \gammav_4 {\cal A}' ) x +  \gammav_5 {\cal A} +  \gammav_6 \dot {\cal A} =0 \;,
\ee
where the coefficients $\gammav_i$ are related to the original EFT functions \eqref{alphabeta}. Given what discussed above, the fact that $x$ in this equation always  appears without derivatives is not surprising because the theory is degenerate and the  scalar degree of freedom must satisfy second-order equations of motion.

Equation \eqref{NLeq} is a cubic polynomial. We search for three real solutions close to the source where the overdensity is large,   ${\cal A} \gg 1$,  that can be matched to the linear unscreened solution away from the source.
If $\gammav_2>0$ \cite{Kimura:2011dc}, two branches of solutions can be obtained for
\be
\label{approx}
x^2  \approx   - \gammav_3 {\cal A} -  \gammav_4 {\cal A}'  \qquad ({\cal A} \gg 1) \;,
\ee
which can be used to solve the equations for $\Phi$ and $\Psi$ for where $\pi$ is nonlinear. 
One obtains (setting $a=1$)
\be
\begin{split}
\label{potentials}
\Phi' (r) &=G_{\rm N} \left( \frac{ m(r)}{r^2} + \Upsour_1 { m''(r)} \right)\;, \\
\Psi' (r) & =G_{\rm N} \left( \frac{ m(r)}{r^2} + \Upsour_2 \frac{  m'(r)}{ r} + \Upsour_3 {  m''(r)} \right)\;.
\end{split}
\ee
These expressions  give the two gravitational potentials close to the matter source, in the regime of large scalar field nonlinearities.  Outside the matter source $m' = m'' =0$ and one recovers the  Newtonian behaviour, $\Phi = \Psi = G_{\rm N} m/r$, although the coupling constant $G_{\rm N}$ is in general time dependent and affected by $\alphaV$ and $\beta_1$. 

To parameterize this possible deviation from standard gravity in screened regions, we introduce the 
fractional difference between the gravitational wave coupling constant $(8 \pi M^2 )^{-1}$ and the effective Newton constant $ G_{\rm N}$, 
\be
\Upsour_0 \equiv (8 \pi M^2 G_{\rm N} )^{-1} -1  = \alphaV - 3 \beta_1\;.
\ee
Inside the matter source, the gravitational potentials are in general different and receive corrections that depend on the density profile of the source (and its radial derivative), similarly to what happens in beyond Horndeski theories \cite{Kobayashi:2014ida}. The corrections are proportional to three time-dependent functions parameterizing the breaking of the Vainshtein screening inside matter,  which can be expressed in terms of the parameters $c_T^2$, $\alphaH$, $\alphaV$ and $\beta_1$ as
\be
\begin{split}
\label{Upsgen}
\Upsour_1 & \equiv \frac{(\alphaH + c_T^2 \beta_1)^2}{c_T^2 (1 + \alphaV - 4 \beta_1 ) -\alphaH -1 } \;, \\
\Upsour_2 & \equiv - \frac{\alphaH (\alphaH - \alphaV +2 (1+ c_T^2 ) \beta_1) + \beta_1  (c_T^2-1) (1+c_T^2 \beta_1)}{c_T^2 (1 + \alphaV - 4 \beta_1 ) -\alphaH  -1} \;, \\
\Upsour_3 & \equiv - \frac{\beta_1 (\alphaH + c_T^2 \beta_1)}{c_T^2 (1 + \alphaV - 4 \beta_1 ) -\alphaH -1 } \;.
\end{split}
\ee
The above expressions, derived  here for the first time, are the 
most general for scalar-tensor theories propagating a single scalar degree of freedom.

The above solutions are not unique. If the right-hand side of eq.~\eqref{approx} is positive (or negative) and $\gammav_2<0$ (or $\gammav_2>0$), there is a  third branch of solutions that can be matched to the linear unscreened solution. This corresponds to taking ${\cal A} \gg x^2 \gg 1$ in eq.~\eqref{NLeq} and the two potentials are obtained by solving eqs.~\eqref{VPhi} and \eqref{VPsi} for $x=0$. One finds  $\Phi'  =G_{ {\rm N},\Phi}  { m}/{r^2}$ and $\Psi'  =G_{ {\rm N},\Psi}  { m}/{r^2}$ with 
\be
\label{thirdbranch}
G_{ {\rm N},\Phi} = \frac{c_T^2}{8 \pi M^2 (1+\alphaH + c_T^2 \beta_1 )^2} \;, \qquad
G_{ {\rm N},\Psi} = \frac{1+\alphaH}{c_T^2} G_{ {\rm N},\Phi}  \;.
\ee
In this case $\Phi \neq \Psi$ even outside the matter source \cite{Kimura:2011dc,Kobayashi:2014ida}, which can be used to place stringent constraints on the free parameters. Because of that, we will not discuss this branch in what follows but we will come back to it  before the conclusion.

\section{Beyond Horndeski theories} 

We now focus on specific cases and compare with results previously found in the literature. Let us  specialize eq.~\eqref{Upsgen} to the beyond-Horndeski (or GLPV) theories, which do not contain higher derivatives in the EFT action \eqref{EFTaction}, i.e.~$\beta_1=\beta_2=\beta_3=0$. In this case,  the expressions for the $\Upsour_I$ simplify, i.e.~$\Upsour_0=1+\alphaV$,
\be
\begin{split}
\label{UpsBH}
\Upsour_1 & = \frac{\alphaH^2}{c_T^2(1 + \alphaV  ) -\alphaH-1 } \;, \quad
\Upsour_2  =  - \frac{\alphaH ( \alphaH - \alphaV) }{c_T^2(1 + \alphaV  ) -\alphaH-1 } \;, 
\end{split}
\ee
and $\Upsour_3 =0$. 

These equations extend the expressions obtained in \cite{Saito:2015fza,Koyama:2015oma} under the assumption that $Q=f_{,\phi}=0$,  in the notation of eq.~\eqref{TypeIa}. The expressions in those references are analogous 
to the ones above but with $\alphaB$ replacing $\alphaV$.
 At first, it is surprising that $\alphaB$ appears in those expressions because, in contrast with $\alphaV$,  in the quasi-static limit
the  operator proportional to $\alphaB$ does not contain  terms quartic in the perturbations---such as the last two terms in eq.~\eqref{Lagrangian}---and  that hence contribute to the Vainshtein mechanism. The explanation is that when $Q= f_{,\phi}=\beta_1=0$
one sees from eq.~\eqref{alphabeta}  that $\alphaV = \alphaB$, so that these expressions can also be  written in terms of $\alphaB$. However,  in general $\Upsour_1$ and $\Upsour_2$ are independent of 
$\alphaB$ and one needs to go beyond the quadratic action
and introduce the dependence on $\alphaV$.

\section{After GW170817}  

The simultaneous observation of GW170817 and GRB 170817A  implies that gravitational waves travel at the speed of light, with very small deviations \cite{Monitor:2017mdv}. This has  dramatically constrained the parameter space of available theories.
To avoid that even $10^{-5}$ fluctuations in the matter overdensity and gravitational potentials along the path of the gravitons affect their speed of propagation, some of the coefficients of the EFT action must be simply set to zero.

It is straightforward to derive these consequences using eq.~\eqref{EFTaction}. The coefficient $c_T^2-1$ detunes the space kinetic term of the gravitons, $(\partial_k \gamma_{ij})^2$, contained in ${}^{(3)}\!R$,
from their
time kinetic term, $\dot \gamma_{ij}^2$, contained in $\delta {\cal K}_2$. Thus,  one must require $c_T=1$. This must be true also for small fluctuations around the background solution. Small fluctuations of the background induce small changes in $\delta N$ in front of $\delta {\cal K}_2$ and ${}^{(3)}\!R$, which modify the  speed of gravitons  if the two coefficients $\alphaH$ and $-\alphaV$ do not coincide. Therefore, gravitons travel at the same speed as photons independently of small changes in the background if  \cite{Creminelli:2017sry}
\be
\label{ctunocond}
c_T=1 \;, \qquad \alphaV = - \alphaH \; . 
\ee
(In eq.~\eqref{alphabeta}, this translates into $a_2 = a_{2,X}=0$.)
Remarkably,  these conditions are stable under quantum corrections \cite{Pirtskhalava:2015nla,Creminelli:2017sry}.

Using these conditions in eq.~\eqref{Upsgen} we find $
\Upsour_0 =  1-\alphaH - 3 \beta_1 $
and 
\be
\Upsour_1 = - \frac{(\alphaH +  \beta_1)^2}{2 (\alphaH + 2 \beta_1)} \;, \quad \Upsour_2 = \alphaH \;, \quad \Upsour_3 = - \frac{\beta_1 (\alphaH +  \beta_1)}{2 (\alphaH + 2 \beta_1)} \;.
\ee
These expressions, also obtained in \cite{Crisostomi:2017lbg,Langlois:2017dyl},  are in general valid independently of $\alphaB$. 
For  $\beta_1=0$ and $Q=f_{,\phi}=0$, they agree with \cite{Sakstein:2017xjx}  where  $\alphaB = - \alphaH$ and become $\Upsour_1 = -\alphaH/2$, $ \Upsour_2 = \alphaH$ and $\Upsour_3=0$.

Before discussing the observational constraints on these expressions, we note that the second condition in eq.~\eqref{ctunocond} does not necessarily apply if dark energy has a fixed $\dot \phi$ independent of $H$, in which case small changes around the background do not  induce a change in $\delta N$ \cite{Creminelli:2017sry}, and if (dark) matter is not coupled to the same metric as photons \cite{Cusin:2017mzw}. In these cases one must use the general expressions in eq.~\eqref{Upsgen}.

\begin{figure}
\begin{center}
\includegraphics[width=8.5cm]{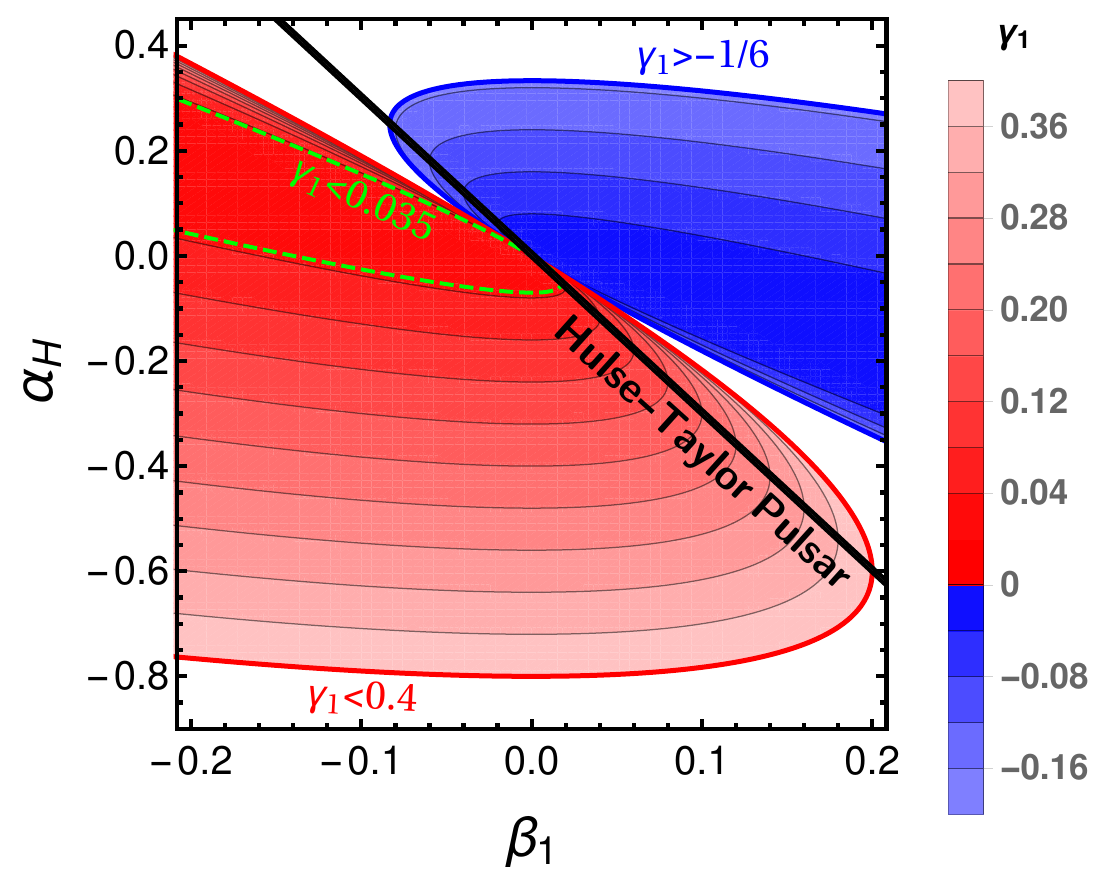}  
\caption{The allowed regions in the plane $(\beta_1,\alphaH)$ after the GW170817 event,   as a function of the upper and lower bound on $\Upsour_1$, are respectively shown in red and blue. We consider only values in the range $-1/6 < \Upsour_1 < 0.4$, to satisfy the stellar structure  \cite{Saito:2015fza} and  minimal mass red dwarf (v3 of \cite{Sakstein:2015zoa})  constraints. (The bound $\Upsour_1 < 0.035$ comes from \cite{Saltas:2018mxc}, see the {\em Note added} below.) The black band represents the Hulse-Taylor  pulsar constraint, $-7.5 \times 10^{-3} \le \Upsour_0-1 \le 2.5 \times 10^{-3} $ \cite{Jimenez:2015bwa}. The region allowed by combining the three observations is given by the overlap of the  black band and the red {\em or} the blue region. } \label{Fig}
\end{center}
\end{figure}

\section{Observational constraints} 

Several late-time observational bounds have been put on the parameters $\Upsour_I$. The Newtonian potential $\Phi$ controls the stellar structure equation, so one can bound $\Upsour_1$ independently of $\Psi$. Since the modification of $\Phi$ are not new, the bounds in the literature straightforwardly apply. In particular, a negative value of $\Upsour_1$ means stronger gravity inside a star so that, for stars to exist in hydrostatic equilibrium, one requires $\Upsour_1 > -1/6$ \cite{Saito:2015fza}. An upper bound, $\Upsour_1 < 0.4$, comes from requiring that the smallest observed red dwarf star has a mass larger than the minimum mass allowing hydrogen to burn in stars, see v3 of Ref.~\cite{Sakstein:2015zoa}. Constraints on $\Upsour_{2}$ and $\Upsour_{3}$ must be derived altogether and require observations involving the curvature potential $\Psi$ and we do not discuss them here.

Let us turn now to $\Upsour_0$, i.e.~the ratio between the screened effective Newton constant, $G_{\rm N}$, and the effective coupling constant for gravitons, $M^{-2}$. 
As shown in Ref.~\cite{Jimenez:2015bwa}, the decrease of the orbital period of binary stars is proportional to $(M^2 G_{\rm N} c_T)^{-1}$. With $c_T=1$, $\Upsour_0$ can be constrained by the 40 year-long observation of the Hulse-Taylor pulsar (PSR B1913+16) \cite{Hulse:1974eb}. Using the results of \cite{Jimenez:2015bwa} based on \cite{Weisberg:2010zz}, one obtains $-7.5 \times 10^{-3} \le \Upsour_0-1 \le 2.5 \times 10^{-3} $  at 2$\sigma$.  
This constraint  assumes that the scalar radiation does not participate to the energy loss. For cubic screening, the effect has been shown to be suppressed by  $-3/2$ powers of the product of the orbital period and the Vainshtein radius \cite{deRham:2012fw,Chu:2012kz}.

As shown in Fig.~\ref{Fig}, combining these constraints  places  tight bounds on $\alphaH$ and $\beta_1$: only the tiny overlap between the black band from the the Hulse-Taylor pulsar 
and the blue and red regions survives,  which implies that
$- 0.60 \le \alphaH \le 0.26$ and $-0.08 \le \beta_1 \le 0.20$. 
For GLPV theories ($\beta_{1}=0$), this leads to a very stringent bound on 
$\alphaH$: $-2.5 \times 10^{-3} \le \alphaH \le 7.5 \times 10^{-3} $ at 2$\sigma$. 
Since the constraints $\Upsour_1$ are likely to improve in the future (see e.g.~the {\em Note added} below), in Fig.~\ref{Fig} we also show the contours corresponding to smaller values of $|\Upsour_1|$.

Let us go back to the third branch of solutions, discussed near eq.~\eqref{thirdbranch}. In this case, the constraints on $\alphaH$ and $\beta_1$ are even more stringent. Indeed, one can use the bound on 
the combination $\Psi/\Phi-1$ from Cassini spacecraft experiment \cite{Bertotti:2003rm}, which for $c_T=1$ translates into $-2.5 \times 10^{-5} < \alphaH  < 6.7 \times 10^{-5}$ (2$\sigma$).  Using this result, the bound on  $(8 \pi G_{{\rm N},\Phi} M^2)^{-1}  = (1+\alphaH + \beta_1)^2 $ from the pulsar discussed above gives $-3.8 \times 10^{-3} < \beta_1< 1.4 \times 10^{-3}$.

\section{Conclusions} 

We have 
obtained the most general expression of the gravitational potentials in the Vainshtein regime, for degenerate higher-order scalar-tensor theories up to quadratic in  second derivatives of the scalar. After GW170817, these are the most general Lorentz-invariant theories propagating a single scalar degree of freedom.
To do so, we have employed the EFT of dark energy approach at nonlinear order and computed  the deviations from  general relativity, outside and inside matter sources. In general, these modifications imply four observational parameters and depend on four EFT parameters, but if gravitons travel at the same speed as photons independently of small changes in the background, they depend only on $\alphaH$ and $\beta_1$, which measure the beyond-Horndeski ``character'' of the theories, see Table~\ref{table}. 
Using these results and those from astrophysical observations, we have obtained stringent constraints on these two parameters.
Our bounds have been derived using only $\Upsour_0$ and $\Upsour_1$. Any  constraints on $\Upsour_{2}$ and $\Upsour_{3}$ will exclude a new region of the $(\beta_1,\alphaH)$ plane, possibly ruling out theories beyond Horndeski.

\section*{Acknowledgements} We thank D.~Langlois, K.~Noui and  F.~Piazza and especially P.~Creminelli  for enlightening discussions and  suggestions and K.~Koyama, J.~Sakstein and M.~Vallisneri for useful correspondences. We also thank M.~Crisostomi for pointing out a (inconsequential) mistake in the first relation in eq.~\eqref{alphabeta}. Moreover, we thank N.~Bartolo, P.~Karmakar, S.~Matarrese and M.~Scomparin for kindly sharing the draft of their paper \cite{Bartolo:2017ibw} before submission. A.D.~acknowledges kind hospitality at IPhT during the completion of this work. F.V.~acknowledges financial support from ``Programme National de Cosmologie and Galaxies'' (PNCG) of CNRS/INSU, France and  the French Agence Nationale de la Recherche under Grant ANR-12-BS05-0002.  We thank  J.~Sakstein and I.~Sawicki for useful correspondences on the point in the ``Note added''.

\vskip.1cm
\emph{Note added.---} An improved upper bound on $\Upsour_1$ was derived  in Ref.~\cite{Saltas:2018mxc}, after the first arXiv submission  of our article. The authors of this reference find $\Upsour_1 < 0.035$, which combined with the other bounds discussed here implies  stringent constraints: $- 0.05 \le \alphaH \le 0.26$ and $-0.08 \le \beta_1 \le 0.02$ at 2$\sigma$, see  Fig.~\ref{Fig}.

\appendix


 \bibliographystyle{utphys}
\bibliography{EFT_DE_biblio3}

\end{document}